# US Corporate Bond Yield Spread
# A default risk debate

Syed Noaman SHAH[*]
*University of Orleans*
s.noamanshah@gmail.com

Mazen KEBEWAR[*]
*University of Orleans*
*University of Aleppo*
mazen.kebewar@univ-orleans.fr

## Abstract

According to theoretical models of valuing risky corporate securities, risk of default is primary component in overall yield spread. However, sizable empirical literature considers it otherwise by giving more importance to non-default risk factors. Current study empirically attempts to provide relative solution to this conundrum by presuming that problem lies in the subjective empirical treatment of default risk. By using post-hoc estimator approach of Lubotsky & Wittenberg (2006), we construct an efficient indicator for risk of default, by using sample of 252 US non-financial corporate data (2000-2010). On average, our results validate that almost 48% of change in yield spread is explained by default risk especially in recent financial crisis period (2007-2009). Hence, our results relatively suggest that potential problem lies in the ad-hoc measurement methods used in existing empirical literature.

**Keywords:** Default risk, credit spread, risk-aversion, measurement error, index construction.

**JEL Classification:** C1, C30, G12, G14.

[*] University of Orleans (France) – Faculty of Law, Economics and Management – Orleans Economic Laboratory (LEO), UMR CNRS 7322, Rue de Blois, B.P. 26739 – 45067 Orléans Cedex 2 – France.

## 1. Introduction

Recent financial crisis has led regulators to increase their attention towards controlling credit risk[1] component of overall corporate structure of financial and non-financial institutions (Schuermann 2004). In addition, yield spread provides support in evaluating whether potential and existing investors are adequately compensated for the risk they are bearing or not, particularly in period of poorer economic outlook which in turn correlated with likelihood of lower-wealth situations invigorating reduce willingness to bear and hold risk by risk-averse investors in capital market.

This aporia lead theoretical financial economists[2] to devise efficient methods to adequately value risky nature of corporate securities. The way theoretical models price these corporate debt securities depend fundamentally upon their related credit risk. On the contrary, extensive empirical literature[3] on the issue finds it practically arduous to accept this conjecture. The purpose of this paper is to attempt relatively to fill the gap by conjecturing that problem lies in ad-hoc measurement techniques employed by existing empirical literature.

The main attribute of current empirical study is to efficiently minimize measurement errors caused due to ad-hoc treatment of default risk in existing empirical literature. By objectively considering risk of default as linear combination of significant proxy factors, current work through post-hoc estimator approach of Lubotsky and Wittenberg (2006) relatively provide solution to this problem.

The rest of paper is organized as follows. Next section delineates briefly related literature and formulates default risk proxy factors. Section 3 outlines estimation methodology and data used. Empirical result analysis is presented in Section 4. Finally, Section 5 sums up the main findings and concludes current work.

## 2. Literature Review

Extensive literature exists on the issue of explaining probable components of corporate bond yield spread. Historically, related literature divides them into two main categories; risk of default and residual risk premium attributing to non-default factors (Fisher 1959). But with the advent of option pricing method of Black and Scholes (1973), new horizon emerges to value risky corporate liabilities. Following Black and Scholes (1973), Merton (1974) gives contingent claim theory by evaluating credit risk of corporate debt as a function of call option on its equity securities. In his seminal work on pricing of risky corporate debt, Merton (1974) mainly focuses to incorporate corporate debt valuation process on the basis of risk of default and non-stochastic interest rate. Later, Black and Cox (1976), Longstaff and Schwartz (1995), among others[4], also treat risk of default as their primary pillar in deriving new valuation models incorporating and relaxing assumptions[5] made by Merton (1974) to value corporate securities.

---

[1] We will use terms of credit risk, default risk and risk of default interchangeably in this paper.

[2] Black and Scholes (1973), Merton (1974), Black and Cox (1976), among others.

[3] Elton et al. (2001), Collin-Dufresne et al. (2001), Delianedis and Geske (2001), Huang & Huang (2003), Tsuji (2005), among others.

[4] Chance (1990), Leland and Toft (1996), Duffie (1999), Longstaff et al. (2005), Ericsson and Renault (2006), Couderc et al. (2007), Liu et al. (2009).

[5] Merton (1974) assumes firm will default only when they fully exhaust all their asset value and include constant interest rate in his valuation model.

Despite immense efforts made by theoretical authors in treating risk of default as primary focus for valuing risky corporate securities, most empirical work do not confirm their premise to give key weight to default risk as the major component of yield spread. Elton et al. (2001), a pioneer empirical work in disentangling the determinants of corporate yield spread show that expected default loss (risk of default proxy) does not contribute to overall yield spread more than 25 percent. Among many others, Jones et al. (1984), Collin-Dufresne et al. (2001), Delianedis and Geske (2001), Huang and Huang (2003), Tsuji (2005), Liu et al. (2009), also show that, in general, risk of default does not contribute majorly in explaining overall yield spread as compared to other non-default factors.

Lack of consensus, between theoretical and empirical proponents, in treating risk of default as a major determinant in yield spread spur the question: Why risk of default is not being proven empirically as a major component in explaining yield spread?.

We argue, in this paper, that one of the main concerns for investor is the risky nature of security for which they ought to be satisfied through the inclusion of default risk premium with other non-default related return, i.e. 'greater the security's risk of default, the greater its default premium should be' (Sharpe et al. 1999: 447). Hence, we conjecture that this controversial issue of empirically lacking to accept risk of default as a major proponent of yield spread is mainly due to the measurement error induced by the usage of different nature of ad-hoc proxies in existing empirical literature. Therefore, we hypothesize that investor's perception of risk of default is a function of more than ad-hoc proxies utilized in previous empirical work.

To serve this purpose, we treat yield spread as a linear function of optimal explicit factors of default risk and control for the factors associated with overall macroeconomic and financial change. In other words, we focus on to build an optimal index for default risk on the basis of these proxies. This in turn enables to assist in solving the subjective treatment of default risk in overall US corporate bond yield spread.

## 2.1 Default Risk Proxies

By convention there is no standard definition of what constitutes a risk of default (Schuermann 2004). Most of the time in existing literature risk of default is being treated on ad-hoc basis. In order to streamline the optimal effect of default risk, we follow the Bank for International Settlements definition of default event (BIS, 2001: 30)[6]. In particular, what we are disposing here is that in order to simplify the conundrum of subjective treatment of default risk, we should treat risk of default as an optimal composite indicator of underlying explicit proxies, which leads to show relative true significance of default risk in overall formulation of yield spread. Bearing in mind the above discussion, we delineate explicit proxy factors of default risk (derived from BIS definition) alongside the control variables.

### 2.1.1 Change in Distance to Default (ΔDD)

This proxy takes into consideration net worth of US corporates and volatility in its stock price. We calculate distance to default as a difference between natural logarithm of total assets and total liabilities divided by its monthly stock price volatility. It, ideally, represents the percentage change in overall long term default probability of the debt issuer (i.e. inability to fulfill its long term obligations). We expect an inverse relation between distance to default and the overall yield spread.

### 2.1.2 Change in Z-score (ΔZ)

Z-score approach predicts corporate probability of default on short term basis by using firm-specific performance ratios and plausibly caters financial health of the obligor past due 90 days or more on any of its credit obligation (Altman 1968). We presume that a positive percentage increase in Z-score expects to lower the corporate bond yield spread.

---

[6] Basel Committee on Banking Supervision 2001.

### 2.1.3 Change in Credit ratings (ΔlR)

In similar vein, credit rating variations help to cater investors' immediate decision in investing and financing activities. We include credit rating proxy because it presents rating agencies' appraisal of risk of default associated with related corporate fixed income security issues. To represent credit rating changes, we use change in downgrade credit rating percentage because it is suitable to assess the impact of credit rating change on overall risk of default. In addition, increase in downgrade credit rating change will certainly impact positively on the overall change in yield spread for corporates'.

### 2.1.4 Cash Flow volatility (CFv)

On the other hand to evaluate firms' position from cash flow perspective, we include fluctuations in operational cash flow situation of debt issuer to assess their ability to service outstanding debt as and when it becomes due. We expect a positive relationship between firm cash flow volatility and the risk-averse investor expected return from the risky corporate security. Following, Tang and Yan (2010), cash flow volatility is calculated as the coefficient of variation of operating cash flows by dividing monthly standard deviation to its absolute mean. In particular, we expect that this proxy will cater the true effect of firm's ability to meet its financial obligation in real terms.

## 2.2 Control variables

In order to evaluate the true effect of change in default risk proxy factors on change in yield spread, we control for different proxy variables reflecting the US macroeconomic and financial market condition. Following Tsuji (2005), as volatility in economic activity leads to affect investors' overall perception of required risk premium, we use change in Industrial Production Index rate (ΔIPI)[7] as a control proxy for change in business cycle (Collin-Dufresne et al 2001). Furthermore, to control for the effects of inflation level, we use change in Consumer Price Index (ΔCPI).[8] We expect a negative impact of change in inflation level on the explained variable (Merton 1974); this negative relationship is also confirmed by Longstaff & Schwartz (1995) and Collin-Dufresne et al. (2001).

On the other hand, to control for financial market indicators, we use change in Fed Fund Rate (ΔFFR) to gauge the inter-bank market effect. In addition, we control for last month change in yield spread on its current month value ($\Delta YS_{t-1}$) following Van Landschoot (2004). We expect a positive relation between these values.

## 3. Data and Methodology

### 3.1 Data

In order to measure optimal effect of risk of default on overall yield spread, monthly data on variable of interest (i.e. default risk proxies) and control variables, outlined in previous section, has been used for the period of 2000 to 2010. This period is suitable for the study as it cited occurrence of two recent financial crises (dot com and subprime crises) in the US economy. For control variables and corporate bond yield spread, entire data is collected from online source of Federal Reserve Bank of St. Louis. On the other hand, for default risk proxy indicators we use Thomson Reuter database to calculate distance to default, cash flow volatility and z-score, individually, for 252 US non-financial corporates. In particular, using New York Stock Exchange (NYSE) Bond Master service we came to include the sample of 252 non-financial US corporates listed with plain vanilla bonds of medium and long term maturities (greater than one year till twenty years). In addition, downgrade percentage change in their credit ratings data is being collected from Standard & Poor's research report 2011.

---

[7] We also use change in GDP (ΔGDP), instead of change in industrial production index (ΔIPI) in order to evaluate robustness and statistical significance of default risk factors.

[8] We also use change in producer price index (ΔPPI) instead of change in consumer price index (ΔCPI) for robustness.

Further, to collect stock prices (daily and monthly) for US corporate's in order to calculate default risk proxy factors, we use Yahoo! Finance website. Additionally, in order to obtain average monthly time series of these explicit proxy factors we include simple weighted average across individual firms'.

In addition, downgrade credit ratings data is available in annual frequency with US GDP data in quarterly frequency. We use cubic spline method to convert data into monthly frequency in order to make these time series in harmony with the rest of data variables. Since, our data follow time series we checked whether it has a unit root or not. Following Phillips-Perron test of unit root (1988), it is observed that almost all of the time series were stationary integrated with zero order. In order to condense the effect of cubic spline algorithm we, in our preliminary model, use natural logarithmic form of change in downgrade credit ratings variable and natural logarithmic form of change in GDP control variable. Further, for monthly yield spread data, we use US treasury monthly yield data (averaged on 1, 2, 5, 10 and 20 years) and US corporate bond monthly yield data (averaged on investment and non-investment grades of similar maturity) from the online source of Federal Reserve Bank of St. Louis. Finally, this gives us total of 132 monthly observations of the focused time series data.

## 3.2  Methodology

Following our premise of objective treatment of default risk, we follow "post-hoc" estimator approach of Lubotsky and Wittenberg (2006). Using multiple regressions (least square) of explanatory proxy variables on dependent variable, Lubotsky and Wittenberg (2006) use regression coefficients of each explanatory proxy to compute a weighted average index, showing true effects of variable of interest. By assuming risk of default as our focused variable, we aim to evaluate the optimal effect of explicit default risk proxy factors on change in the relative yield spread while controlling for macroeconomic and financial market variables.

Since, weights used here are ratio of covariances of dependent variable (ΔYS) and statistically significant default risk proxy variables to the covariance between the most significant proxy variable[9] and dependent variable, this technique helps in optimally minimizing variance of errors terms of these proxy variables that leads to construct our required composite indicator. Thus, it enables us to reduce noise or measurement error caused, in the existing empirical literature, due to using biased coefficients of ad-hoc proxies treated as the risk of default.

Hence, we use this statistic as an optimal indicator reflecting the true effect of risk of default by utilizing previously explained proxies simultaneously through multiple regressions. Therefore, we first evaluate, individually, effects of our default risk proxy factors on change in yield spread and then following our principal objective, evaluate linearly effect of our default risk proxy factors, aggregately, on change in overall yield spread.

This can be manifested as;

$$\Delta YS_t = \theta + \alpha \Delta X_t + \gamma \Delta W_t + \xi_t \tag{1}$$

---

[9] We follow Akaike information criteria (AIC) to select the principal proxy factor.

Whereas, ($\theta$) represents constant term, ($\Delta X_t$) represents matrix (vector) of monthly change in default risk factor time series (t-1 to t), ($\Delta W_t$) represents matrix (vector) of change in control variable time series from month t-1 to month t, ($\alpha$) is a coefficient matrix (scalar) of our explanatory proxy variables, ($\gamma$) is a coefficient matrix (scalar) of control variables, ($\xi_t$) shows error term following i.i.d, and ($\Delta YS_t$) represents change in overall yield spread (i.e. difference between US corporate bond yield—non financial and US treasury yield, on average, month t-1 to month t of similar maturity).

We can describe the final form of our model as;[10]

$$\Delta YS_t = \theta + \alpha_1 CFv_t + \alpha_2 \Delta DD_t + \alpha_3 \Delta Z_t + \alpha_4 \Delta lR_t + \gamma_1 CPI_t + \gamma_2 \Delta IPI_t + \gamma_3 \Delta FFR_t + \gamma_4 \Delta YS_{t-1} + \xi_t \quad (2)$$

Whereas, ($\theta$) represents constant term, ($\Delta$) shows monthly percentage change from month t-1 to month t, ($\alpha$) represents coefficients of explanatory proxy variables, ($\gamma$) shows control variables coefficients, ($\xi_t$) is a disturbance term following i.i.d and ($\Delta YS_t$) represents change in overall yield spread.

Following, Lubotsky and Wittenberg (2006), we treat risk of default as a composite explanatory variable which is not been optimally measured in existing empirical literature. Thus it is being represented by an optimal mix of significant proxy factors. Stated differently, to show the optimal effect of risk of default factor on change in yield spread of non-financial US corporate securities, we use the post-hoc estimator approach.

In addition, our presumption delineates the objective treatment of risk of default based on proxy factors derived from BIS definition of default event. On the basis of equation (2), we include those proxy variables which are statistically significant in order to construct our representative indicator for risk of default.

$$\rho_t = \sum_{j=1}^{k} \frac{Cov(Y_t, X_{jt})}{Cov(Y_t, X_{1t})} * \beta_j \quad (3)^{11}$$

In equation (3), ($\rho_t$) shows our post-hoc estimator for the risk of default indicator in month t. Further, ($Y_t$) represents our dependent variable (change in yield spread—$\Delta YS$) in month t, and ($X_{jt}$) represents our statistically significant proxy factors j in month t, whereas ($\beta_j$) represents the least square regression coefficient of significant (j) explanatory variables at time t (months). Further ($X_{1t}$) presents principal statistically significant explanatory variable[12] on which the post-hoc estimator is constructed.

Equation (3) gives us the optimal proxy indicator for default risk by minimizing measurement errors in order to evaluate its true effect on the overall yield spread formulation for risky corporate securities. To assess this effect, we regress US corporate bond yield spread on our post-hoc risk of default estimator.

$$\Delta YS_t = \phi + \lambda \Delta \rho_t + v_t \quad (4)$$

In equation (4), ($\Delta YS_t$) represents difference between US corporate bond yield (investment and non-investment grade) and the US treasury yield, on average, in month t. ($\rho_t$) represents the default risk post-hoc estimator in month t, whereas ($\lambda$) shows the regression coefficient of our default risk estimator and ($v_t$) represents i.i.d error term. $\Delta$ represents the monthly percentage change in variables (from month t-1 to month t). Further, ($\phi$) shows the constant term in equation (4). In order to evaluate robustness of our default risk optimal indicator, we take into consideration its effect on overall yield spread during recent US financial crisis (July 2007 to March 2009). We divide our sample time interval into crisis and non-crisis period and analyze our explanatory variable effect on dependent variable.[13] For this we introduce dummy variable into specification (4) and our estimated model becomes;

---

[10] Following Collin-Dufresne et al. (2001) empirical model.

[11] Detail derivation can be found in Lubotsky and Wittenberg (2006).

[12] We follow Akaike information criteria (AIC) to select the principal factor as $X_{1t}$.

$$\Delta YS_t = \phi + \lambda_1 \Delta \rho_{t\ (crisis)} + \lambda_2 \Delta \rho_{t\ (non-crisis)} + v_t \tag{5}$$

In equation (5), $\Delta \rho_{t\ (crisis)}$ reports change in default risk post-hoc during financial crisis and $\Delta \rho_{t\ (non-crisis)}$ shows change in default risk post-hoc during non-crisis period.

Specification (5) mainly explains how change in our risk of default indicator affects change in overall yield spread during crisis and non-crisis period. We use interaction term of our focused estimator with the crisis and non-crisis dummy variable in order to effectively gauge the behavior of our explanatory variable on dependent variable during the crisis period. Next section delineates and analyzes the econometric results and limitations of specifications introduced in this section.

## 4. Empirical Analysis

Our preliminary objective is to study monthly percentage change of the default risk explanatory proxies on our dependent variable (i.e. change in yield spread), therefore, focused time series become stationary integrated with order zero after taking percentage change effect.[14]

Since our explanatory proxy variable time series are stationary, we follow ordinary least square (OLS) regression for the econometric estimation. Table (1), column (1) shows the result of introducing cash flow volatility as a risk of default proxy to explain changes in yield spread (while controlling for macroeconomic and financial market factors). In column (1), the operational cash flow volatility (CFv) of non-financial US firms significantly affect change in relative yield spread following presumed relation as delineates in section 2.1.4.

In particular, if we glance on overall results in all columns of table (1), then it is evident that all our explanatory proxy variables follow expected relationship with our dependent variable. Furthermore, in column (2), (3) and (4) we evaluate, individually, the effects of change in distance to default (ΔDD), change in Z-score (ΔZ) and change in downward grade credit rating (ΔlR) on our explained variable (ΔYS). We find that (ΔDD) affects positively and (ΔlR) affects negatively our dependent variable (ΔYS) at 1% significant level, respectively. In addition, (ΔZ) although shows expected relation with dependent variable is not significant. On the other hand, if we look at our control variables in column (1) to (4), we find that change in inter-bank interest rate and change in previous period yield spread (month t-1) significantly affects change in current period yield spread (month t) following presumed relation (as explained in section 2.2).

Table (1), column (5), report results of our specification (2), which delineates the linear effect of change in default risk proxy factors on change in overall yield spread, while catering the effects of control variables. After reviewing the results of first four columns of table (1), we may be factual to validate our conjecture of treating risk of default as a linear combination of these proxy factors as their explanatory power significantly enhances to (37%) in rationalizing overall variation in US corporate bond yield spreads.

---

[13] We deeply thank anonymous reporter comments regarding taking into consideration recent US financial crisis disruptions in our analysis.

[14] We use Augmented Dickey Fuller and Philips Perron tests of stationarity to check the presence of unit roots. All our time series are stationary integrated with order zero i.e. I (0).

Following the first four columns, we evaluate the results of our main model (column 5, table 1). In column (5), we discover that cash flow volatility (CFv), change in distance to default (ΔDD) and change in downgrade credit rating (ΔlR) remains statistically significant. On the contrary, change in Z-score (ΔZ) although reports presumed relation with change in yield spread (ΔYS) remains statistically insignificant.

**Table 1**: Influence of change in risk of default proxy factors on US Corporate bond yield spread change ($\Delta YS_t$)

|  | 1 | 2 | 3 | 4 | 5 | 6 |
|---|---|---|---|---|---|---|
| CFv | 0.74$^c$ |  |  |  | 1.51$^c$ | 1.30$^c$ |
|  | (0.37) |  |  |  | (0.92) | (0.79) |
| ΔDD |  | -0.15$^a$ |  |  | -0.15$^a$ | -0.17$^a$ |
|  |  | (0.03) |  |  | (0.03) | (0.03) |
| ΔZ |  |  | -0.07 |  | -0.04 | -0.02 |
|  |  |  | (0.05) |  | (0.03) | (0.03) |
| ΔlR |  |  |  | 0.90$^b$ | 1.17$^a$ | 1.21$^a$ |
|  |  |  |  | (0.45) | (0.42) | (0.49) |
| Δ CPI | -0.15 | -1.97 | -0.63 | -1.89 | -4.65$^b$ |  |
|  | (2.02) | (1.92) | (1.99) | (2.21) | (2.09) |  |
| Δ IPI | -0.09 | -0.43 | -0.31 | -0.01 | -0.65 |  |
|  | (1.02) | (0.95) | (1.01) | (1.00) | (0.92) |  |
| ΔFFR | -3.24$^a$ | -2.55$^a$ | -3.18$^a$ | -3.22$^a$ | -2.54$^a$ | -2.62$^b$ |
|  | (0.56) | (0.54) | (0.55) | (0.55) | (0.52) | (1.56) |
| $\Delta YS_{t-1}$ | 1.47$^a$ | 1.24$^a$ | 1.62$^a$ | 1.57$^a$ | 1.34$^a$ | 1.25$^b$ |
|  | (0.55) | (0.51) | (0.54) | (0.54) | (0.50) | (0.53) |
| ΔPPI |  |  |  |  |  | -1.82$^a$ |
|  |  |  |  |  |  | (0.61) |
| ΔlGDP |  |  |  |  |  | -1.79 |
|  |  |  |  |  |  | (1.20) |
| Observations | 131 | 131 | 131 | 131 | 131 | 131 |
| R² Adjusted | 0.22 | 0.2 | 0.15 | 0.24 | 0.37 | 0.30 |
| D-W | 1.79 | 1.91 | 1.88 | 1.83 | 1.99 | 1.97 |
| LM- Statistics (B-G test) |  |  |  |  | 15.50 | 16.97 |
| P-value (B-G test) |  |  |  |  | 0.21 | 0.15 |

**Note:** (CFv) Cash Flow volatility, (ΔDD) change in Distance to Default, (ΔZ) change in Z-score, (ΔlR) change in Credit Ratings, (ΔCPI) change in Consumer Price Index, (ΔIPI) change in Industrial Production Index rate, (ΔFFR) change in Fed fund Rate, ($\Delta YS_{t-1}$) change in previous period Yield Spread, (ΔPPI) change in Producer Price Index, (ΔlGDP) change in GDP. Standard errors are reported in brackets. a, b & c represent significance level at 1%, 5% & 10% respectively.

Thus, results in column (5) table (1), describe significant long term effect of default risk proxy factors on the dependent variable. In general, one percent increase in distance to default for US non financial firms (i.e. showing that a firm is moving away from the level of default) reduces related yield spread percentage by almost 0.15 units. Further, a one percentage point increase in operational cash flow volatility increases vulnerability of technical insolvency of US non-financial firms and hence raises their respective yield spread by almost 1.51units. In similar vein, a percentage increase in downgrade credit rating increases the overall yield spread by almost 1.2 units, validating presence of risk-averse investors in capital market.

On the other hand, with the linear introduction of default risk proxy factors, we find that change in inflation (ΔCPI) becomes statistically significant. Whereas, change in interest rate (ΔFFR) effect and change in last month yield spread ($\Delta YS_{t-1}$) remains statistically significant, respectively.[15] In particular, our results confirm the relationship between inflation and interest rate on overall yield spread put forward by theoretical strand of related literature. Furthermore, Breusch-Godfrey Lagrange Multiplier test (B-G test) suggests that we cannot reject null hypothesis of no serial correlation in residual error terms (i.e. p-value of 0.21).

---
[15] As constant term (θ) is statistically insignificant with minimal regression coefficient, we drop related results.

Column (6), table (1), for robustness, show the results which are not significantly different from results of column (5).[16] In addition, B-G test for column (6) also reports no evidence of significant serial correlation in our residual error terms.

On the other hand, we did not find any significant correlation between default risk proxy factors that indicate presence of any multicollinearity problem. Furthermore, correlations between these proxy variables follow, in general, their expected signs. But in order to be certain that results reported in table (1) are not exacerbated by this problem, we also follow the variance decomposition proportions method of detecting significant multicollinearity among the explanatory variables (Belsley et al. 1980) and the variance inflation factor (VIF). From the analysis of correlation matrix, variance decomposition proposition method and VIF indicator, we deduce that our proxy factors of default risk do not possess the problem of significant multicollinearity which leads us to evaluate the results of our specification (3), (4) and (5) respectively.[17]

Following specification (3), from table (1) column (5), we took only those proxy factors which show statistically significant relation and construct a monthly time series of change in default risk factor delineating the optimal mix effect of statistically significant proxies. In particular, we select[18] distance to default as our principal proxy factor ($X1_t$ in specification 3) representing optimal default risk indicator.

Figure 1; graphically depict change in default risk post-hoc indicator time series ($\Delta\rho_t$) along with change in yield spread ($\Delta YS$) of US corporate bonds for period of 2000-2010. It shows that our risk of default estimator almost systematically depicts a structural relation with overall yield spread of non-financial US corporate bond securities, showing similar long term trend. Further, our variable of interest ($\Delta\rho_t$) quite effectively gauge the dot-com (2000-2002) and sub-prime (2007-2009) crises, reporting more than 320 basis points change in the overall risk of default during sub-prime crisis and almost 200 basis points increase in the overall risk of default for US Corporate bond issues in dot-com crisis.

**Figure 1:** Default risk post-hoc estimators ($\Delta\rho_t$) and yield spread of US corporate bond securities ($\Delta YS$)

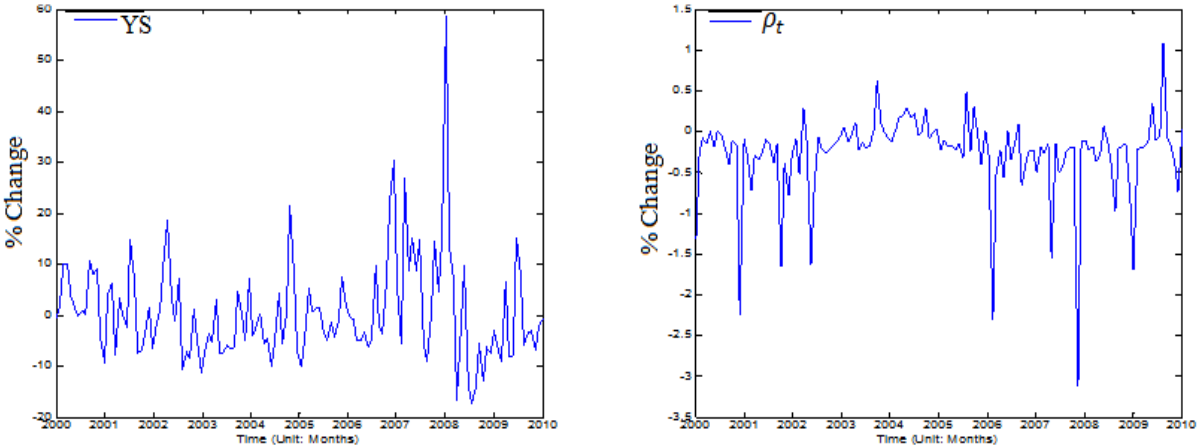

---

[16] We replace control variables, further, we also introduce, simultaneously, change in stock market return ($\Delta SP$), S&P500, and change in its volatility index ($\Delta VIX$) in our focused model, but results did not significantly varies.

[17] Related results could be provided upon request.

[18] On the basis of Akaike Information Criteria (AIC).

Table (2), column (1) show the results of our specification (4), which is the focused objective of this study i.e. evaluating the effect of optimal risk of default estimator on overall yield spread of US corporate bonds.

Column (1) table (2), shows that our post-hoc default risk estimator is statistically significant in explaining variations in US corporate bond yield spread. In particular, regression coefficient indicates that due to one percentage positive change in default risk estimator (i.e. as it is based on distance to default principal proxy factor, therefore it means firm is moving away from default level) the respective yield spread reduces by (0.62) units. In addition, explanatory power of our risk of default indicator is (41%) which relatively supports theoretical model assumptions in treating risk of default as a primary component in valuing risky corporate securities.[19] Furthermore, significant constant term in specification (4) shows presence of non-default factors in overall formulation of corporate bond yield spread. On the other hand B-G test show no evidence of serial correlation in residual error terms. In addition, through Ramsey Reset test, we validate the evidence of delineating specifications based on our surmise of objectively treating default risk as linear combination of significant proxy variables.[20] In turn, these results support existing empirical literature of linear relationship between focused dependent and explanatory variables.

**Table 2:** Default risk post-hoc estimator effect on US corporate bond yield spread

| Dependent variable is yield spread change ($\Delta YS_t$) | 1 | 2 |
|---|---|---|
| $\Delta \rho_t$ | -0.62[c] | |
|  | (0.33) | |
| $\Delta \rho_{t\ (crisis)}$ | | -1.22[a] |
|  | | (0.46) |
| $\Delta \rho_{t\ (non-crisis)}$ | | -0.50 |
|  | | (0.71) |
| $\phi$ | 0.04[c] | 0.64[b] |
|  | (0.02) | (0.33) |
| Observations | 132 | 132 |
| R² Adjusted | 0.41 | 0.48 |
| D-W | 1.99 | 2.05 |
| LM-Statistics (B-G) | 5.90 | 1.94 |
| P-value (B-G) | 0.93 | 0.37 |
| F-statistics (Ramsey Reset Test) | 0.06 | 0.76 |
| P-value (Ramsey Reset Test) | 0.93 | 0.46 |

**Note**: [$\Delta \rho_t$] change in default risk post-hoc, [$\Delta \rho_{t\ (crisis)}$] change in default risk post-hoc during financial crisis, [$\Delta \rho_{t\ (non-crisis)}$] change in default risk post-hoc during non-crisis period. Standard errors are reported in brackets. a, b & c represents significance level at 1%, 5% & 10% respectively.

In addition, table (2) column (2) report results of specification (5) which delineates the effect of financial system disruptions in the US due to sub-prime crisis (2007-2009). In particular, results in column (2) outline how significantly our focused optimal indicator of default risk affects yield spread variations in this crisis period. It is really interesting to note that by bifurcating our focused time interval into crisis and non-crisis period, our risk of default estimator shows engaging effect on overall fluctuation in yield spread. It now manifest that statistically significant relation in column (1) is mainly due to the effects of financial crisis of 2007-2009. Table (3) column (2) shows that our default risk estimator is statistically significant during sub-prime crisis period in explaining variations in US corporate bond yield spread.

---
[19] Supporting Churm & Panigirtzoglou (2005) and Tang & Yan (2010).
[20] Here, p-value is more than 5%, so we can not reject null hypothesis of correct linear model specification.

On the contrary, during normal (i.e. non-crisis) period it is not statistically significant though following presumed relation with our dependent variable. In addition, significance level of our constant term, table (2) column (2), has increased emphasizing the importance of presence of non-default risk factors in overall formulation of US corporate bond yield spread in this crisis period.[21] Further, explanatory power of our post-hoc estimator increased by 7 percent (i.e. to 48%), indicating robust effects in explaining abrupt variations in overall yield spread in sub-prime crisis period. In similar vein, results of Breusch Godfrey LM test of serial correlation for residual error term shows no evidence of significant serial correlation. In general, by comparing results in column (1) and (2), table (2), we draw potential outcome that our post-hoc estimator mainly gauge significant relation on overall yield spread in the sub-prime crisis period and its effect dilates across non-crisis period on aggregate basis. This in turn relatively supports theoretical strand of related literature in treating default risk as a primary pillar in the valuation of risky corporate securities.[22]

## 5. Conclusion

The aim of this study is to explore the treatment of default risk conundrum between theoretical and empirical proponents of valuing risky corporate debt securities. The dominant view exists in theoretical literature is that risk of default is fundamental to value risky corporate securities which is evident in structural models of valuing credit spread. Whereas major empirical literature on the issue shows that non-default risk factors are more important in valuing these risky corporate securities. Current study attempts to fill this void by presuming that problem to this debate lies in the measurement techniques existing empirical literature use as a subjective treatment of default risk. Using risk of default as an objective linear function of significant proxy factors, we construct an optimal indicator (following Lubotsky and Wittenberg 2006) for default risk which contributes relatively to solve this debate. Our risk of default estimator significantly explains the change in US corporate bond yield spread during recent financial crisis (2007-2009). In particular, it explain about (48%) change in overall yield spread model, indicating dominance of theoretical[23] school of thought in treating default risk as a cornerstone in risky corporate debt valuation process. Further, it will be interesting to include sample from other developed markets (such as European Union) and from emerging markets to validate robustness of our results.

---

[21] Supporting results of Dick-Nielsen et al. (2012).

[22] We also run regressions by introducing dot-com crisis (2000-2002) effect but results were not statistically significant probably showing limitations of our sample that lacks data in this sector.

[23] Provided by Merton (1974).